\documentclass[amsmath,amssymb]{revtex4}
\usepackage{bm}
\usepackage{amsthm}
\usepackage{amsmath}
\usepackage{amsfonts}
\usepackage{amssymb}

\theoremstyle{plain}

\theoremstyle{definition}

\usepackage[]{graphicx}
 
\begin{document}

\title{Nonlocal Transformations for Accelerated Observers}

\author{Bahram Mashhoon}
\email{mashhoonb@missouri.edu}
            
\affiliation{Department of Physics and Astronomy\\University of Missouri\\Columbia, Missouri 65211, USA}

\begin{abstract}
 According to the locality postulate of special relativity, the measurements of physical fields by accelerated observers at a given event in Minkowski spacetime are related to each other by the representations of the Lorentz group. Nonlocal extensions of these representations are necessary, however, once acceleration-induced nonlocality is taken into account. The particular case of Dirac spinors is treated in detail and the corresponding nonlocal transformation group is studied.
\end{abstract}

\maketitle                   
  
\section{Introduction}\label{sec:1}

Imagine a global inertial frame with coordinates $x^\mu =(t,\boldsymbol{X})$ in Minkowski spacetime and the set of ideal inertial observers at rest in this frame. The basic non-gravitational laws of physics have been formulated with respect to these \textit{fundamental} observers. Each such observer carries the orthonormal tetrad frame $\bar{\lambda}^\mu _{\;\;(\alpha)}=\delta^\mu_{\;\;\alpha}$ along its worldline. With respect to the fundamental observers, the electromagnetic field, for instance, has components $F_{\mu\nu}(x)$ that satisfy Maxwell's equations of electrodynamics.

All real observers are more or less accelerated. 
At an arbitrary event $P$, consider the set of all possible observers at $P$. This set includes accelerated observers as well as ideal inertial observers. It is a consequence of the special theory of relativity that a field according to any member of this set at $P$ is related to the field according to the fundamental observer at $P$ by a (matrix) representation of the Lorentz group. This assertion is based on the hypothesis of locality, which relates accelerated observers with ideal inertial observers. Special relativity assumes that an accelerated observer is at each instant locally inertial; that is, it is physically equivalent to an otherwise identical momentarily comoving inertial observer whose worldline is tangent to the worldline of the accelerated observer at that instant. This locality postulate is extended in nonlocal special relativity by including a certain average over the past worldline of the accelerated observer~\cite{1}. The purpose of this paper is to show that in this case the representation of the Lorentz group should be extended to a nonlocal representation. That is, at each event $P$ the field measurements of different accelerated observers are related to each other by nonlocal representations of the Lorentz group. This will be illustrated in this paper in detail for the Dirac spinor; a similar treatment applies to other fields.

The motivation for nonlocal special relativity has been discussed at length in a number of publications---see~\cite{1} and references therein. It is based on a detailed analysis of field measurements by accelerated  observers. If all physical phenomena could be reduced to \textit{pointlike coincidences}, then the standard special relativity would be completely adequate. Indeed, nonlocal special relativity reduces to the standard theory when the intrinsic scale of the phenomenon under observation is negligible compared to the acceleration scales of the observer. The need for acceleration-induced nonlocality may be expressed via the Bohr-Rosenfeld principle~\cite{2,3,4}.

Bohr and Rosenfeld~\cite{2,3} discussed electromagnetic measurements of the fundamental observers and pointed out a dichotomy regarding the physical interpretation of the equations of classical electrodynamics. The electromagnetic field component $F_{\mu \nu}(x)$ at an event $x$ is physically meaningful only in terms of an average over a sufficiently small spacetime region $\Delta$ about the event $x$,
\begin{equation}\label{eq:1} \langle F_{\mu\nu}\rangle =\frac{1}{\Delta} \int_\Delta F_{\mu\nu}(x')d^4x',\end{equation}
since the measurement of the component of the field by the fundamental observers around $x$ using the Lorentz force law cannot be done at an event but requires extended measurements in space and time around $x$. Nevertheless, experience shows that the predictions of the \textit{local} differential field equations involving $F_{\mu\nu}(x)$ generally agree with the results of experiments involving \textit{averages} of $F_{\mu\nu}(x)$. That is, to the extent that observers can be identified in practice with the fundamental inertial observers, agreement can be achieved between the predictions of \textit{local} field equations with \textit{nonlocal} experimental results.

The basic analysis of Bohr and Rosenfeld~\cite{2,3} simply ignored the fact that all actual observers are accelerated. It is true that in many low-acceleration situations of interest, the effects due to the observers' acceleration are rather small and can be safely neglected~\cite{1}. As a matter of principle, however, it is necessary to take the acceleration of the observers into account. In this case, the accelerated observers' local tetrad frames $\lambda^\mu_{\;\;(\alpha )} (x)$ become additional variables that enter the averaging process in~\eqref{eq:1}. To ensure that the predictions of the theory still correspond to observational results, the nonlocality involved in the averaging process must become an integral part of the basic theory. That is, the electromagnetic field component $\mathcal{F}_{\mu\nu}(x)$ according to accelerated observers would still be local, but would satisfy appropriate integro-differential equations. The precise form of such equations and their physical implications are the subject of nonlocal special relativity~\cite{1,4}.

The plan of this paper is as follows. The nonlocal Dirac spinor is discussed in section~\ref{sec:2}. The corresponding nonlocal representation of the Lorentz group and its properties are worked out in section~\ref{sec:3}. It is demonstrated explicitly via an example in section~\ref{sec:4} that, in contrast to the local representation of the Lorentz group, the nonlocal version is not unitary. Section~\ref{sec:5} contains a discussion of our results.

\section{Nonlocal Dirac spinor}\label{sec:2}

According to the fundamental inertial observers---i.e. those at rest---in the background global inertial frame under consideration in this paper, a free Dirac spinor $\psi (x)$ satisfies the Dirac equation
\begin{equation}\label{eq:2} (i\gamma^\alpha \partial _\alpha -m)\psi (x)=0,\end{equation}
where $m$ is the mass of the Dirac particle and $\gamma^\alpha$ are the Dirac matrices. Here we follow the standard conventions of~\cite{5} and set $c$ and $\hbar$ equal to unity.

Consider next an accelerated observer following a path $x^\mu(\tau )$ in spacetime, where $\tau$ is its proper time. We define the orthonormal tetrad frame $\lambda^\mu_{\;\;(\alpha)}(\tau )$ along this worldline via~\cite{1,4}
\begin{equation}\label{eq:3} \frac{d\lambda^\mu_{\;\;(\alpha)}}{d\tau }=\Phi _{(\alpha)}^{\;\;\;\;(\beta)}\lambda^\mu_{\;\;(\beta )},\end{equation}
where $\Phi_{(\alpha )(\beta)}$ is the antisymmetric acceleration tensor. According to the locality postulate, the accelerated observer may in effect be replaced by a continuous sequence of otherwise identical momentarily comoving inertial observers. The comoving inertial observer's tetrad frame at $\tau$ is thus given by $\lambda^\mu_{\;\;(\alpha)}(\tau)$ as well; moreover, its Dirac spinor can be expressed as
\begin{equation}\label{eq:4} \hat{\psi} (\tau)=\Lambda (\tau )\psi (\tau ),\end{equation}
where~\cite{6,7,8}
\begin{equation}\label{eq:5} \Lambda (\tau )=e^{-\int^\tau_{\tau_0}k(\tau ')d\tau '}\Lambda (\tau _0)\end{equation}
and $k(\tau)$ is given by
\begin{equation}\label{eq:6} k(\tau )=\frac{i}{4}\Phi_{(\alpha)(\beta)}(\tau )\sigma^{\alpha\beta}.\end{equation}
Here $\tau_0$ is the instant at which the acceleration is turned on and $\sigma^{\alpha\beta}$ is defined by~\cite{5}
\begin{equation}\label{eq:7} \sigma^{\alpha\beta}=\frac{i}{2}[\gamma^\alpha ,\gamma^\beta].\end{equation}

Under an inhomogeneous Lorentz transformation $x\to x'$, where $x^\alpha =L^\alpha_{\;\;\beta} x^{\prime\beta}+s^\alpha$, the Dirac spinor transforms as~\cite{5}
\begin{equation}\label{eq:8} \psi '(x')=S\psi (x),\end{equation}
where $S$ is the spin transformation matrix such that
\begin{equation}\label{eq:9} S\gamma^\alpha S^{-1}=L^\alpha _{\;\;\beta }\gamma^\beta .\end{equation}
The locality postulate involves instantaneous inhomogeneous Lorentz transformations from the global inertial frame to the local inertial frame of the momentarily comoving inertial observer. Thus imagine a continuous series of transformations of the form~\eqref{eq:8}-\eqref{eq:9}. The tetrad frame is constructed from the basis vectors of the instantaneous inertial rest frames of the momentarily comoving observers. Hence under the Lorentz transformation, the corresponding tetrads transform as
\begin{equation}\label{eq:10} \bar{\lambda}^\mu _{\;\;(\alpha)} L^\alpha_{\;\;\beta}=\lambda^\mu_{\;\;(\beta)},\end{equation}
so that with $\bar{\lambda}^\mu_{\;\;(\alpha)}=\delta^\mu_{\;\;\alpha}$, we have $L^\alpha_{\;\;\beta}=\lambda^\alpha_{\;\;(\beta)}$. Therefore, the corresponding spin transformation matrix at instant $\tau$ is $\Lambda (\tau)$ such that
\begin{equation}\label{eq:11} \Lambda \gamma^\alpha \Lambda^{-1}=\lambda^\alpha_{\;\;(\beta)}\gamma^\beta.\end{equation}
It is straightforward to show that~\eqref{eq:4}-\eqref{eq:7} are compatible with~\eqref{eq:11}; in fact, $S\to \Lambda$ and $\psi '(x')\to \hat{\psi}$. It follows from these considerations that $\Lambda$ is a $4\times 4$ matrix representation of the Lorentz group.

Let us next turn to nonlocal special relativity, where the Dirac spinor according to the accelerated observer is given for $\tau \geq \tau_0$ by~\cite{1}
\begin{equation}\label{eq:12} \hat{\Psi} (\tau )=\hat{\psi} (\tau )+\int^\tau_{\tau_0} k(\tau ')\hat{\psi} (\tau ')d\tau '.\end{equation}
The nonlocal part of this relation, which can be neglected in the eikonal limit, has the form of a certain average over the past worldline of the observer. According to Volterra's theorem, the relationship between $\psi (\tau )$ and $\hat{\Psi}(\tau )$ is unique in the space of continuous functions~\cite{9}; Volterra's uniqueness result has been extended to the Hilbert space of square-integrable functions by Tricomi~\cite{10}.

It is necessary to extend~\eqref{eq:12} to a congruence of accelerated observers. To avoid unphysical situations, we assume that the acceleration of the congruence is turned on at a certain time $t_i$ and turned off at a later time $t_f$. In fact, as discussed in detail in~\cite{11}, we assume that the observers under consideration are confined to an open bounded spacetime domain $\Omega$ in which
\begin{equation}\label{eq:13} \hat{\Psi}(x)=\hat{\psi}(x) +\int_\Omega \hat{K} (x,x')\hat{\psi} (x')d^4x'\end{equation}
together with $\hat{\psi}(x)=\Lambda (x)\psi (x)$ constitutes a generalization of~\eqref{eq:12} to the congruence such that the relation between $\hat{\Psi}(x)$ and $\psi (x)$ is unique. The explicit construction of the kernel $\hat{K}(x,x')$ for special cases has been discussed in~\cite{1,11}. We assume that $\Omega$ is large enough to contain all possible congruences of accelerated observers under consideration here.

The integral equation~\eqref{eq:13} has the unique solution 
\begin{equation}\label{eq:14} \hat{\psi} (x)=\hat{\Psi} (x)+\int_\Omega \hat{R}(x,x')\hat{\Psi} (x') d^4x'\end{equation}
via the Liouville-Neumann method of successive substitutions. In fact, using this method, \eqref{eq:13} can be written as a uniformly convergent series
\begin{equation}\begin{split}\label{eq:15} \hat{\psi} (x) &= \hat{\Psi} (x)-\int_{\Omega} \hat{K} (x,y)\hat{\Psi} (y)d^4y\\
&\quad +\int_\Omega \int_\Omega \hat{K} (x,y)\hat{K}(y,z)\hat{\Psi} (z)d^4yd^4z-\dots .\end{split}\end{equation}
It is useful to define iterated kernels $\kappa_n,n=1,2,\dots ,$ by
\begin{equation}\label{eq:16} \kappa_1 (x,y) =-\hat{K} (x,y),\quad \kappa_{n+1}(x,y)=\int_\Omega \kappa_1 (x,z)\kappa_n(z,y)d^4z.\end{equation}
These kernels satisfy the relation $(p=1,2,\dots)$
\begin{equation}\label{eq:17} \kappa_{n+p}(x,y)=\int_\Omega \kappa _n(x,z)\kappa_p(z,y)d^4z.\end{equation}
Moreover, the reciprocal kernel $\hat{R}(x,y)$ can be expressed as
\begin{equation}\label{eq:18} \hat{R} (x,y)=\sum^\infty_{n=1}\kappa_n(x,y).\end{equation}
Let us note here two important relations connecting the reciprocal kernels that simply follow from~\eqref{eq:16}-\eqref{eq:18}, namely,
\begin{equation}\begin{split}\label{eq:19} \hat{K}(x,y)+\hat{R}(x,y)&=- \int_\Omega \hat{K} (x,z)\hat{R}(z,y)d^4z\\
&=- \int_\Omega \hat{R} (x,z)\hat{K} (z,y)d^4z.\end{split}\end{equation}
The comparison of~\eqref{eq:13} with~\eqref{eq:14} makes it evident that $\hat{K}$ is reciprocal to $\hat{R}$ as well; moreover, substituting~\eqref{eq:14} in~\eqref{eq:13} and vice versa demonstrate that relations~\eqref{eq:19} are indeed necessary for the sake of consistency.

Using these results, it is possible to express $\psi (x)=\Lambda^{-1}\hat{\psi} (x)$ in terms of $\hat{\Psi}$ via~\eqref{eq:14}, so that the nonlocal Dirac equation becomes
\begin{equation}\label{eq:20} (i\gamma^\alpha\partial_\alpha -m)\Lambda^{-1}(x)\left[\hat{\Psi}(x)+\int_\Omega \hat{R}(x,y)\hat{\Psi}(y)d^4y\right]=0.\end{equation}
As in previous work~\cite{8,11}, one can define
\begin{equation}\label{eq:21} \Lambda^{-1}\hat{\Psi}=\Psi\end{equation}
and
\begin{equation}\label{eq:22} \Lambda^{-1}(x)\hat{R}(x,y)\Lambda (y)=R(x,y),\end{equation}
so that the nonlocal Dirac equation takes the simpler form
\begin{equation}\label{eq:23} (i\gamma^\alpha \partial_\alpha -m)\left[\Psi (x)+\int_\Omega R(x,y)\Psi (y)d^4y\right]=0.\end{equation}

The spinor $\Psi$ has been employed extensively in~\cite{1} in the study of the nonlocal interaction of a charged Dirac particle with the electromagnetic field; however, the equations of motion in~\cite{1} can be simply reformulated in terms of $\hat{\Psi}$ via~\eqref{eq:21}. This latter approach may be preferable due to the direct physical significance of $\hat{\Psi}$.

For the purposes of the present paper, the significance of these considerations lies in the connection between the Dirac spinor according to the accelerated observers at $x$, $\hat{\Psi} (x)$, and the Dirac spinor $\psi (x)$ according to the fundamental inertial observer at $x$; this relation can be written as
\begin{equation}\label{eq:24} \hat{\Psi}(x)=\Lambda (x)\left[\psi (x)+\int_\Omega K(x,y)\psi (y)d^4y\right].\end{equation}
This is obtained from~\eqref{eq:13} via
\begin{equation}\label{eq:25} \Lambda^{-1} (x) \hat{K}(x,y)\Lambda (y)=K(x,y);\end{equation}
moreover, we note from~\eqref{eq:21} that
\begin{equation}\label{eq:26}\Psi (x)=\psi (x)+\int_\Omega K(x,y)\psi (y)d^4y.\end{equation}
Here $K(x,y)$ and $R(x,y)$ are reciprocal kernels; they satisfy relations similar to~\eqref{eq:19}, as can be easily demonstrated by multiplying~\eqref{eq:19} by $\Lambda^{-1}(x)$ from the left side and by $\Lambda (y)$ from the right side.

\section{Nonlocal transformation group}\label{sec:3}

At an event $x$ in spacetime, imagine the set of all accelerated as well as ideal inertial observers at $x$ and their corresponding Dirac spinors. We wish to relate the physical determinations of one observer to another; specifically, we are interested in the transformation of one spinor into another. Let $g$ be a linear operator at $x$,
\begin{equation}\label{eq:27} g[\psi ] (x)=\Lambda (x)\left[\psi (x)+\int_\Omega K(x,y)\psi (y)d^4y\right],\end{equation}
which transforms the spinor of the fundamental observer at $x$ into the spinor of an arbitrary accelerated observer at $x$---see~\eqref{eq:24}. This transformation can be simply expressed as $\hat{\Psi}=g\psi$. We wish to show that the set $G$ of transformations given by~\eqref{eq:27} forms a group under composition. Let us recall here that for a given fixed event $x$, the set of all $\Lambda (x)$ forms a representation of the Lorentz group.

Writing~\eqref{eq:27} formally as $g=(\Lambda ,K)$, we note that $G$ is closed under composition; that is,
\begin{equation}\label{eq:28} g_1g_2=(\Lambda_{12},K_{12}),\end{equation}
where $\Lambda_{12}=\Lambda_1\Lambda_2$ and $K_{12}$ is given by
\begin{equation}\label{eq:29} K_{12} (x,y)=\tilde{K}_1 (x,y)+K_2 (x,y)+\int_\Omega \tilde{K}_1 (x,z)K_2 (z,y)d^4z.\end{equation}
Here
\begin{equation}\label{eq:30} \tilde{K}_1 (x,y)=\Lambda_2^{-1}(x)K_1 (x,y)\Lambda_2 (y).\end{equation}
There is therefore an identity transformation $e=(I,0)$, where $I$ is the unit $4\times 4$ matrix, since $eg=ge=g$. Moreover, every element of the group $g=(\Lambda ,K)$ has an inverse $g^{-1}=(\Lambda^{-1},\hat{R})$, where $\hat{R}$, as defined by~\eqref{eq:14}, is the reciprocal of $\hat{K}$. Using~\eqref{eq:19}, it is straightforward to check that $g^{-1}g=gg^{-1}=e$. It remains to show that associativity holds, namely, $g_1(g_2g_3)=(g_1g_2)g_3$. A detailed calculation shows that this is indeed the case; in fact,
\begin{equation}\label{eq:31} g_1g_2g_3=(\Lambda_1\Lambda_2\Lambda_3,K_{123}),\end{equation}
where $K_{123} $ is given by
\begin{equation}\begin{split}\label{eq:32} K_{123} (x,y)&=K''_1(x,y) +K'_2 (x,y)+K_3 (x,y)\\
&\quad +\int_\Omega K''_1 (x,z)K'_2 (z,y)d^4 z\\
&\quad +\int_\Omega K''_1 (x,z)K_3 (z,y)d^4z\\
&\quad + \int_\Omega K'_2 (x,z)K_3 (z,y)d^4z\\
&\quad +\int_\Omega \int_\Omega K''_1 (x,z)K'_2 (z,w)K_3 (w,y)d^4zd^4 w.\end{split}\end{equation}
Here
\begin{align}\label{eq:33} K''_1 (x,y)&=\Lambda^{-1}_{23} (x)K_1 (x,y)\Lambda_{23} (y),\\
\label{eq:34} K'_2 (x,y)&=\Lambda^{-1}_3 (x)K_2(x,y)\Lambda_3 (y),\end{align}
and $\Lambda_{23}=\Lambda_2\Lambda_3$.

The group elements of the form $\bar{g}=(\Lambda ,0)$ constitute a subgroup $\bar{G}$ of $G$. This is the subgroup of transformations between the spinors associated with ideal inertial observers at $x$; it is indeed a representation of the Lorentz group. For actual accelerated observers, however, $K$ is in general nonzero and we have a nonlocal representation of the Lorentz group. In this connection, let us recall that $K(x,y)$ in~\eqref{eq:27} is given by $\Lambda^{-1}(x)\hat{K}(x,y)\Lambda (y)$, where $\hat{K}(x,y)$ is in turn a generalization of $k=-(d\Lambda /d\tau )\Lambda^{-1}$ in~\eqref{eq:6} to a congruence of accelerated observers in spacetime. Thus for a given congruence, $K$ is essentially determined by $\Lambda$; this means that it is reasonable to regard the transformation group $G$ as a nonlocal representation of the Lorentz group.

 Imagine next the set of all accelerated observers at $x$ for which $\Lambda (x)$ happens to be the identity matrix. That is, these special observers all have tetrad frames at $x$ given by $\lambda ^\mu_{\;\;(\alpha)}(x) =\delta^\mu_{\;\;\alpha}$, which coincides with that of the fundamental observer at $x$.  The corresponding elements of $G$ are of the form   $g^0 = ( I, K )$ and these constitute another subgroup $G^0$ of $G$. 

A significant feature of the nonlocal group $G$ must be mentioned here: The local representation of the Lorentz group $\bar{G}$ is unitary, while the corresponding nonlocal representation is not in general unitary. This is demonstrated in the next section using a simple example involving uniformly rotating observers.

\section{Uniformly rotating observers}\label{sec:4}

Consider an observer that for $t>0$ rotates uniformly with frequency $\omega >0$ on a circle of radius $r\geq 0$ about the $Z$ axis. We assume that for $t<0$, the observer has uniform rectilinear motion parallel to the $Y$ axis such that $X=r$, $Y=r\omega t$ and $Z=Z_0$. At $t=\tau _0=0$, the observer is forced to follow a circle in the $Z=Z_0$ plane such that $X=r\cos \phi$ and $Y=r\sin \phi$, where $\phi=\omega t=\gamma \omega \tau$. The observer's speed is always $\beta =r\omega$ and $\gamma$ is the associated Lorentz factor. Thus for different values of $\omega >0$, $r\geq 0$ and $Z_0$, $-\infty <Z_0<\infty$, we have a whole class of observers uniformly rotating about the $Z$ axis for $t\geq 0$. The orthonormal tetrad frame of such an observer for $t\geq 0$ is given by
\begin{align}\label{eq:35} \lambda^\mu _{\;\;(0)}&=\gamma (1,-\beta \sin \phi ,\beta \cos \phi ,0),\\
\label{eq:36}\lambda^\mu _{\;\;(1)}&=(0,\cos \phi ,\sin \phi ,0),\\
\label{eq:37} \lambda^\mu_{\;\;(2)}&=\gamma (\beta ,-\sin \phi ,\cos \phi ,0),\\
\label{eq:38} \lambda^\mu_{\;\;(3)}&=(0,0,0,1).\end{align}
Using~\eqref{eq:3} and~\eqref{eq:5}-\eqref{eq:7}, it is possible to determine $k(\tau )$ and $\Lambda (\tau )$ for this observer.

Next, we imagine positive-energy plane-wave solutions of the free Dirac equation~\eqref{eq:2} propagating along the positive $Z$ direction with momentum $p$ and spin parallel $(\psi_+)$ or antiparallel $(\psi_-)$ to the $Z$ axis. That is,
\begin{align}\label{eq:39} \psi_\pm &=\chi_\pm e^{-iEt+ipZ},\\
\label{eq:40} \chi_+&=N\begin{bmatrix} 1 \\0\\\rho \\0\end{bmatrix} ,\quad \chi_-=N\begin{bmatrix}0 \\ 1 \\0 \\-\rho\end{bmatrix},\end{align}
where $E=\sqrt{m^2+p^2}$, $p/(m+E)=\rho$, and $N$ is a positive normalization factor. In this case, the calculations of $\hat{\psi}=\Lambda \psi$ and $\hat{\Psi}$ using~\eqref{eq:12} have been carried out in detail in section~III of~\cite{8}. The results are~\cite{8}
\begin{align}\label{eq:41} \hat{\psi}_\pm &=\hat{\chi}_\pm e^{-iE'_\pm \tau+ipZ_0},\\
\label{eq:42} \hat{\chi}_+&=N'\begin{bmatrix} \gamma+1\\-i\beta \gamma \rho\\(\gamma +1)\rho\\-i\beta \gamma\end{bmatrix} ,\quad \hat{\chi}_-=N'\begin{bmatrix}-i\beta \gamma \rho\\\gamma +1\\i\beta \gamma\\ -(\gamma +1)\rho\end{bmatrix},\end{align}
where
\begin{equation}\label{eq:43} E'_\pm =\gamma \left( E\mp \frac{1}{2}\omega \right), \quad N'=\frac{N}{\sqrt{2(\gamma +1)}}.\end{equation}
Moreover,
\begin{equation}\label{eq:44} \hat{\Psi} _\pm =F_\pm (\tau )\hat{\psi}_\pm ,\end{equation}
where
\begin{equation}\label{eq:45} F_\pm (\tau)=1\pm \frac{1}{2}\gamma \omega \frac{1-e^{iE'_\pm \tau}}{E'_\pm}.\end{equation}
The expression for $E'_\pm$ in~\eqref{eq:43} illustrates the phenomenon of spin-rotation coupling---see~\cite{1,4,7} and references therein. A typographical error in the expression for $N'$ in terms of $N$ in the sentence containing equation~\eqref{eq:40} in section~III of~\cite{8} should be corrected: a slash indicating division is missing there.

Under a Lorentz transformation $x\to x'$, the spinor transforms as in~\eqref{eq:8}, while the adjoint spinor transforms as
\begin{equation}\label{eq:46} \bar{\psi}'(x')=\bar{\psi} (x)S^{-1},\end{equation}
so that
\begin{equation}\label{eq:47}\bar{\psi}'(x')\psi'(x')=\bar{\psi}(x)\psi (x)\end{equation}
is a Lorentz scalar. We recall that $\bar{\psi}:=\psi^\dagger\gamma^0$, where $\psi^\dagger$ is the Hemitian conjugate of $\psi$. Equation~\eqref{eq:46} follows from~\eqref{eq:8} and $S^{-1} =\gamma^0S^\dagger\gamma^0$, so that the spin transformation matrix $S$ is not a unitary matrix. These same properties carry over to the matrix $\Lambda$ based on the treatment of section~\ref{sec:2}. In fact, equations~\eqref{eq:4}-\eqref{eq:7}, $\gamma^0 \gamma^{\mu^\dagger} =\gamma^\mu \gamma^0$, and
\begin{equation}\label{eq:48} \gamma^0 \sigma^{\alpha \beta^\dagger}\gamma^0=\sigma ^{\alpha \beta},\quad \gamma ^0 k^\dagger\gamma^0=-k\end{equation}
result in
\begin{equation}\label{eq:49} \Lambda^{-1}=\gamma^0 \Lambda^\dagger \gamma^0.\end{equation}
The matrix $\Lambda$ is a unitary representation of the Lorentz group, since it follows from~\eqref{eq:4} and~\eqref{eq:49} that
\begin{equation}\label{eq:50}\bar{\hat{\psi}} \hat{\psi} =\bar{\psi}\psi.\end{equation}
This equation is consistent with the result of a direct calculation in the case of the uniformly rotating observer, namely,
\begin{equation}\label{eq:51} \bar{\hat{\psi}}_\pm (\tau )\hat{\psi}_\pm (\tau )=\bar{\psi}_\pm \psi_\pm =N^2(1-\rho^2).\end{equation}
On the other hand, the nonlocal spinor given by~\eqref{eq:44} is such that
\begin{equation}\label{eq:52} \bar{\hat{\Psi}}_\pm \hat{\Psi}_\pm =|F_\pm (\tau )|^2 N^2(1-\rho ^2).\end{equation}
It follows from~\eqref{eq:45} that
\begin{equation}\label{eq:53} |F_\pm (\tau )|^2=1\pm 2E\omega \left[ \frac{\gamma \sin \left( \frac{1}{2}E'_\pm \tau \right)}{E'_\pm }\right]^2;\end{equation}
hence,
\begin{equation}\label{eq:54} |F_+(\tau )|\geq 1,\quad |F_-(\tau )|\leq 1.\end{equation}
Thus nonlocality is directly responsible for the lack of unitarity in~\eqref{eq:52}. That is, nonlocality directly affects the wave amplitude as in~\eqref{eq:44}: The amplitude as measured by the rotating observer is higher (lower) if the spin of the incident particle along the axis of rotation of the observer is positive (negative)---i.e. if the particle spin is in the same (opposite) sense as the rotation of the observer. This nonlocal consequence of the coupling of spin with rotation has been further discussed in~\cite{1,4,8,12}.

\section{Discussion}\label{sec:5}

In this paper we consider the set of all possible observers at an event $x$ and their determinations of some physical field at $x$ in accordance with nonlocal special relativity. We are interested in the transformations between these field determinations at $x$, since these form a group that is a nonlocal representation of the Lorentz group. The case of Dirac spinors and the corresponding nonlocal transformation group $G$ is illustrated in detail. For instance, the spinors $\hat{\Psi}_1(x)$ and $\hat{\Psi}_2(x)$ according to observers $1$ and $2$, respectively, are related to each other by
\begin{equation}\label{eq:55} \hat{\Psi}_2 (x)=g_2g_1^{-1} \hat{\Psi }_1(x),\end{equation}
where $g_1$ and $g_2$ are certain elements of the nonlocal group $G$ described in section~\ref{sec:3}. An important aspect of $G$ is that it is not in general unitary, in contrast to the corresponding local subgroup $\bar{G}$ of $G$ that is restricted to ideal inertial observers at $x$ and is a unitary representation of the Lorentz group.

\end{document}